\documentclass[10pt,journal,compsoc]{IEEEtran}

\usepackage{amsmath,amssymb,amsfonts}
\usepackage{makecell,booktabs,tabularx,array}
\usepackage{graphicx}
\usepackage{placeins}
\usepackage{algorithm,algorithmicx,algpseudocode}
\usepackage{cite,url,textcomp,stfloats,balance,xcolor}
\usepackage{tikz}
\usetikzlibrary{arrows.meta,positioning,fit}
\newcommand{\EvalSeeds}{24}

\newcommand{\EvalPlanningMeanMS}{11.42}
\newcommand{\EvalFallbackPct}{86.17}

\newcommand{\EvalCheckpointMB}{0.02}
\newcommand{\EvalNetworkMB}{45.76}

\newcommand{\EvalModelNLL}{0.0217}
\newcommand{\EvalModelDurationError}{0.2002}
\newcommand{\EvalModelBrier}{0.0538}
\newcommand{\EvalModelDisruptionNLL}{0.1966}

\title{\Large Reliability-aware AI Agent Execution in Collaborative Edge Computing:\\ A World Model Approach}
\author{Xiangchun~Chen, Jiannong~Cao,~\IEEEmembership{Fellow,~IEEE,}
Longxiang~Gao, Xiaoming~Wu,
and Yusheng~Ji,~\IEEEmembership{Fellow,~IEEE}%
\IEEEcompsocitemizethanks{%
\IEEEcompsocthanksitem Xiangchun Chen and Jiannong Cao are with the Department of Computing, The Hong Kong Polytechnic University, Hong Kong SAR, China (e-mail: xiang-chun.chen@connect.polyu.hk; jiannong.cao@polyu.edu.hk).
\IEEEcompsocthanksitem Longxiang Gao and Xiaoming Wu are with the Shandong Computer Science Center, Qilu University of Technology (Shandong Academy of Sciences), Jinan, China (e-mail: gaolx@sdas.org; wuxm@sdas.org).
\IEEEcompsocthanksitem Yusheng Ji is with the National Institute of Informatics, Tokyo, Japan (e-mail: kei@nii.ac.jp).
}}

\begin{document}
\IEEEtitleabstractindextext{%
\begin{abstract}

Collaborative edge computing supports AI agent applications by distributing execution across heterogeneous user-controlled devices with limited resources. This paper studies reliability-aware AI agent execution under node failures and wireless disruptions. These disruptions can make agent execution state unavailable or inconsistent, forcing recovery or recomputation that consumes the remaining deadline budget. Maintaining recoverable execution states can therefore improve request completion before deadlines. We formulate an online AI agent execution management problem that jointly selects resource allocations, state protection actions, and recovery targets to maximize the expected task success rate subject to resource, compatibility, and deadline constraints. We develop a world model that jointly captures evolving agent workflows and recoverable execution state under candidate management actions, accounting for node reliability, network conditions, and resource availability. The model enables the controller to simulate future execution and recovery trajectories under candidate actions and estimate their effects on execution progress and task completion before deadlines. We then employ world model-based model predictive control (WorldMPC) with Monte Carlo rollouts to jointly optimize resource allocation, state protection, and recovery decisions for long-term reliable agent execution under the dynamic environment. We evaluate the proposed approach on the simulation and real-world testbed using real-world and synthetic datasets. Experimental results show that WorldMPC improves task success rate over baselines in both simulation and real-world testbed.

\end{abstract}
}

\maketitle
\IEEEdisplaynontitleabstractindextext
\IEEEpeerreviewmaketitle

\section{Introduction}\label{sec:introduction}

AI agents combine language-model reasoning with tool use to complete tasks on user-controlled infrastructure. Collaborative edge computing (CEC) lets heterogeneous devices share compute, memory, models, and tools beyond a single device's capabilities~\cite{borzunov2023petals,zhang2024edgeshard}. Node failures and wireless disruptions make reliable completion before a deadline a central concern.

Reliability is particularly challenging for stateful agent execution. Reasoning outputs can determine subsequent tool operations, while tool outcomes can reveal new subtasks, change dependencies, and increase the remaining workload. Resuming execution requires reasoning context, completed subtask and tool results, generated files, and environment information. A failure can destroy this state, whereas a disconnected link can make surviving state temporarily inaccessible. Restoring this state or reconstructing lost work consumes the remaining deadline budget.  Protecting recent state can reduce recovery work but competes with execution for memory and bandwidth.

Existing studies address agent orchestration~\cite{kim2024an,zhang2025aflow}, inference-state preservation and recovery~\cite{strati2024dejavu,zhang2026efficient,cao2026lumen}, and deadline- and reliability-aware workflow scheduling~\cite{devaraj2026darcs}. The remaining challenge is to coordinate these capabilities when intermediate results change the workflow and failures change its recoverable progress.

We study reliability-aware AI agent execution in collaborative edge computing. Each request follows a dynamically revealed directed acyclic graph (DAG) of dependent reasoning and tool subtasks. A reliability controller jointly determines resource allocation, state protection, and recovery targets under compute, memory, bandwidth, and compatibility constraints. Our objective is to maximize the expected task success rate: the expected fraction of submitted requests whose complete, protocol-consistent outputs reach clients before their deadlines.

Three challenges arise in this problem. First, dynamic dependencies and uncertain reasoning or tool outcomes make the remaining workload and execution time difficult to predict. Second, useful recovery requires consistent state across reasoning progress, tool results, and files. The controller must decide which progress to protect and where, considering both state availability and the work needed to resume execution. Finally, resource allocation, protection, and recovery have coupled effects over time. These effects must be assessed against changing risks and the remaining deadlines of multiple requests.

To address these challenges, we propose a world model-based model predictive control (WorldMPC) approach. Our world model jointly represents dynamic workflow progress and the execution state available for consistent recovery. It predicts how resource allocation, state protection, and recovery decisions affect subsequent execution, state availability, and recovery costs. Learned event predictions and explicit execution rules generate trajectories that respect committed execution history. Building on planning with learned dynamics~\cite{hafner2019learning}, WorldMPC uses Monte Carlo rollouts to compare feasible joint resource, protection, and recovery plans by their estimated on-time task success rate. It executes the first action of the selected plan and replans after receiving new observations.

The main contributions of this work are as follows:

\begin{itemize}
\item \textbf{Joint reliability-aware execution modeling.} We formulate agent execution as a joint resource-allocation, state-protection, and recovery problem. The model links dynamically revealed dependencies and recoverable progress to task deadlines under shared resource and compatibility constraints.

\item \textbf{World model-based planning for reliable agent execution.} We develop a world model that jointly represents dynamic workflow progress and recoverable execution state. It generates execution and recovery trajectories consistent with committed history. WorldMPC uses these trajectories to compare joint resource, protection, and recovery plans by their estimated on-time task success rate.

\item \textbf{Simulation and real-world testbed evaluation.} We compare WorldMPC with state-of-the-art baselines on simulation and real-world testbed using real-world and synthetic datasets. Experimental results demonstrate that our approach improves task success rates over the baselines.

\end{itemize}

The remainder of this paper is organized as follows. Section~\ref{sec:related_work} reviews related work. Section~\ref{system-model-and-problem-formulation} presents the system model and problem formulation. Section~\ref{sec:solution} develops the proposed method. Section~\ref{sec:performance-evaluation} presents the experimental evaluation, and Section~\ref{sec:conclusion} concludes the paper.

\section{Related Work}\label{sec:related_work}

We review related work from three perspectives: AI agent execution and orchestration, fault-tolerant execution and state recovery, and reliability-aware scheduling.

\subsection{AI Agent Execution and Orchestration}\label{subsec:rw_agent_execution}

Yao et al.~\cite{yao2023react} proposed ReAct, which interleaves reasoning, actions, and observations to revise decisions using environmental feedback. However, selecting and using external tools remains a central challenge. Schick et al.~\cite{schick2023toolformer} proposed Toolformer, which learns when to call APIs, what arguments to provide, and how to incorporate their outputs. However, learning to invoke tools does not itself provide a mechanism for retaining lessons from unsuccessful task attempts. Shinn et al.~\cite{shinn2023reflexion} introduced Reflexion to address this issue through linguistic feedback and an episodic memory of reflections. These methods improve tool selection and task-solving behavior within an agent trajectory, while larger collaborative workflows require additional mechanisms.

Wu et al.~\cite{wu2024autogen} developed AutoGen to combine customizable agents and tools through programmable conversations, but developers must specify suitable interaction patterns. Hong et al.~\cite{hong2024metagpt} proposed MetaGPT to structure collaboration through standardized operating procedures, role assignments, and intermediate artifacts. This reduces inconsistencies in software development workflows, but does not automatically determine an efficient order for independent tool calls. Kim et al.~\cite{kim2024an} addressed this issue with LLMCompiler, which separates function-call planning, task fetching, and parallel execution. Parallel dispatch reduces serialization, while selecting reasoning and action trajectories remains a separate planning problem. Zhou et al.~\cite{zhou2024language} developed Language Agent Tree Search, combining Monte Carlo tree search, self-reflection, and environment feedback to explore alternatives. Its search concerns task-solving trajectories rather than device placement or the survival of execution state after infrastructure failures.

Hu et al.~\cite{hu2025automated} introduced Automated Design of Agentic Systems, in which a meta agent programs candidate agents and uses an archive of previous discoveries to guide further search. Zhang et al.~\cite{zhang2025aflow} developed AFlow, which searches over code-represented workflows using Monte Carlo tree search and execution feedback. These methods reduce dependence on manually designed agent structures, but do not eliminate the communication overhead of multi-agent execution.  To address this overhead, Zhang et al.~\cite{zhang2025cut} proposed AgentPrune, which prunes redundant communication in a spatial-temporal message-passing graph.  However,  it does not determine how the remaining computation and state transfers should be scheduled across unreliable local devices. Our focus is the execution progress that these processes produce: committed reasoning outputs, tool results, and artifacts used by later subtasks.

\subsection{Fault-Tolerant Execution and State Recovery}\label{subsec:rw_state_recovery}

Moritz et al.~\cite{moritz2018ray} proposed Ray, which supports dynamic task and actor execution and reconstructs lost task outputs from lineage. However, it cannot guarantee state consistency when tasks modify persistent application state. Zhang et al.~\cite{zhang2020fault} addressed stateful serverless workflows, combining logging, fault tolerance, and transactional execution. Jia et al.~\cite{jia2021boki} developed Boki, which exposes shared logs to support consistent, durable state and fault-tolerant workflows. However, the recovery guarantees of these runtimes do not automatically extend to external tool operations.

For LLM inference, PagedAttention reduces KV-cache fragmentation and supports cache sharing~\cite{kwon2023efficient}, but does not itself retain a surviving copy when a worker becomes unavailable. Llumnix migrates running requests and in-memory state across model instances for load balancing and isolation~\cite{sun2024llumnix}. ServerlessLLM combines checkpoint locality, fast model loading, and live inference migration to reduce startup latency~\cite{fu2024serverlessllm}. However, unexpected source loss requires explicit recovery mechanisms.

D\'{e}j\`{a}Vu uses KV-cache streaming and replication to preserve progress after failures~\cite{strati2024dejavu}. SpotServe adapts model parallelization to preemptible GPU availability and commits inference progress for stateful recovery~\cite{miao2024spotserve}. However, transferring or replicating large inference states can consume scarce bandwidth. EdgeFlow combines direct KV transmission with activation-based recomputation to reduce migration time in collaborative edge inference~\cite{zhang2026efficient}. However, it does not jointly consider migration time and the execution load of resumed requests when selecting recovery destinations. LUMEN jointly considers checkpoint placement, interrupted-request distribution, and capacity restoration during model reload~\cite{cao2026lumen}. 

For agents, committed reasoning outputs can determine later tool operations and dependent subtasks. Recovery must therefore align saved reasoning progress with committed tool outcomes, files, and environment changes. If a repeated model call produces a different output, subsequent execution must still respect the committed history.

\subsection{Reliability-aware Scheduling}\label{subsec:rw_reliability_scheduling}

Qin et al. proposed a multi-objective memetic algorithm for makespan, cost, and reliability in multi-cloud workflows~\cite{qin2023reliability}. However, its backup placements are determined before execution, without adaptive task recovery after failures. RECR addresses energy- and deadline-constrained IoT workflows; RECR-D adds adaptive task duplication to tolerate transient and permanent failures~\cite{taghinezhadniar2025fault}. DARCS formulates deadline- and reliability-constrained admission and scheduling for LLM agentic workflows on heterogeneous edge infrastructure; DARCS-H combines reliability-adjusted priorities, reliability-budget tracking, and constraint repair~\cite{devaraj2026darcs}. However, these studies do not jointly control the protection and restoration of evolving agent execution states as new work emerges from intermediate results.

AlpaServe uses model placement and parallelism to exploit statistical multiplexing under bursty demand~\cite{li2023alpaserve}. However, it does not account for the distinct resource demands of prefill and decoding. DistServe separates prefill and decoding and optimizes their resource allocations and parallelism~\cite{zhong2024distserve}, while Sarathi-Serve uses chunked prefills and stall-free scheduling to balance throughput and latency~\cite{agrawal2024taming}. However, these systems focus on request-level serving rather than dependencies across application workflows. Parrot addresses this limitation by exposing application dataflow through semantic variables to optimize dependent LLM requests~\cite{lin2024parrot}. However, it does not jointly consider execution-state protection and failure recovery in its scheduling decisions.

TetriSched combines runtime estimates and deadlines with repeated replanning under changing cluster conditions~\cite{tumanov2016tetrisched}. However, it does not model performance dependencies across application components. Sinan learns microservice dependencies to allocate resources while maintaining end-to-end tail-latency targets~\cite{zhang2021sinan}. For multi-stage serverless workflows, AQUATOPE uses Bayesian models for container prewarming and resource allocation to support workflow-level QoS under uncertainty~\cite{zhou2023aquatope}. However, these systems do not jointly consider predictive resource allocation, execution-state protection, and failure recovery.

As agent workflows evolve and failures change recoverable progress, protecting a committed cut can reduce repeated reasoning and tool work, but copying and restoring it consume resources shared with ongoing execution. This coupling motivates joint resource allocation, state protection, and recovery to maximize the expected task success rate under node and link disruptions.

\section{System Model and Problem Formulation}\label{system-model-and-problem-formulation}

In this section, we formulate reliability-aware AI agent execution as a joint online problem of advancing an evolving workflow and preserving its recoverable state. Reasoning and tool outcomes both reveal new work and create state dependencies. We model how execution, protection, and recovery share resources and jointly determine completion before deadlines. Table~\ref{tab:main_notations} summarizes the main notations.

\begin{table}[t]
    \caption{Main Notations Used in the Paper}
    \centering
    \label{tab:main_notations}
    \footnotesize
    \renewcommand{\arraystretch}{1.12}
    \begin{tabularx}{\columnwidth}{@{}p{0.29\columnwidth}@{\hspace{0.7em}}X@{}}
        \toprule
        Notation & Description \\
        \midrule
        $\mathcal{N},\mathcal{E}$ & Computing nodes and communication links \\
        $a_i,d_i,D_i$ & Arrival, target latency, and absolute deadline \\
        $\Omega_i$ & Required model and tool set of request $i$ \\
        $\mathcal{A}_i(t)$ & Currently revealed agent workflow DAG \\
        $\mathcal{V}_i(t),\mathcal{E}_i(t)$ & Revealed subtasks and dependency edges \\
        $W_i(t)$ & Remaining workload of request $i$ \\
        $\mathcal{C}_{iv}(t)$ & Compatible targets for subtask $v$ of request $i$ \\
        $p_{iv}(t),x_{ivn}(t)$ & Authoritative executor and binary active assignment \\
        $n_i(t)$ & Number of committed execution events \\
        $\mathcal{O}_i(k),V_i(k)$ & State-object manifest and logical bytes at cut $k$ \\
        $k_{in}^{\mathrm{bk}}(t)$ & Largest complete compatible acknowledged cut retained at node $n$ \\
        $\delta_{in}(t)$ & Committed-event lag of the retained checkpoint \\
        $\mathcal{B}_{in}(t),\Delta V_{in}$ & Validated destination objects and incremental copy bytes \\
        $\mathcal{K}_{in}^{\mathrm{avail}}(t)$ & Complete compatible cuts accessible to recovery target $n$ \\
        $\kappa_i(t)$ & Selected recovery cut; zero denotes initial reconstruction subject to replay feasibility \\
        $G_n^r(t),\gamma_{ivn}^r(t)$ & Compute capacity and subtask allocation in pool $r$ \\
        $\gamma_n^{r,\mathrm{mgmt}}(t)$ & Compute demand of protection and recovery \\
        $M_{nh},W_{nh}$ & Storage capacity and resident model-weight occupancy \\
        $\mathcal{Q}_{nh}(t)$ & Distinct allocated physical state objects in tier $h$ \\
        $M_{nh}^{\mathrm{other}}(t)$ & Workspace, staging, and background occupancy \\
        $\mathcal{F}(t),\mathcal{P}_f$ & Active flows and the path of flow $f$ \\
        $B_e(t),u_f(t),w_f(t)$ & Link capacity, goodput, and remaining transfer bytes \\
        $X_t,U_t,\Xi_t$ & System state, management action, and disturbances or newly observed agent outcomes \\
        $H_t,\Pi_{\mathrm{adm}}$ & Available history and admissible nonanticipative policies \\
        $\mathcal{R},N$ & Submitted-request cohort and its size \\
        $C_i^\pi,Y_i^\pi$ & Protocol-valid client-delivery time and on-time indicator \\
        $Z^\pi,J(\pi)$ & Cohort task success rate and its expectation \\
        $S_k,\sigma_k$ & Latent state and observed state representation at decision epoch $k$ \\
        $\widehat{p}_{\phi},\phi$ & Learned event predictor and its parameters \\
        $\Pi,\widehat{J}_k(\Pi)$ & Candidate feedback plan and its estimated on-time delivery reward \\
        $\mathcal{I}_k,\widehat{C}_{i,m}^{\Pi}$ & Requests with unresolved on-time outcomes and valid delivery time in rollout $m$ \\
        $B,K$ & Caps on candidate-generation attempts and inspected action tuples \\
        $H,M,L$ & Optimized-prefix steps, rollouts per plan, and continuation-event cap \\
        $\Omega_k,\mu_{\mathrm{base}}$ & Planning time budget and continuation/fallback policy \\
        \bottomrule
    \end{tabularx}
\end{table}

\subsection{Agent Execution Model}\label{subsec:sys_system}\label{subsec:sys_task}

\begin{figure*}[t]
    \centering
    \includegraphics[width=\textwidth]{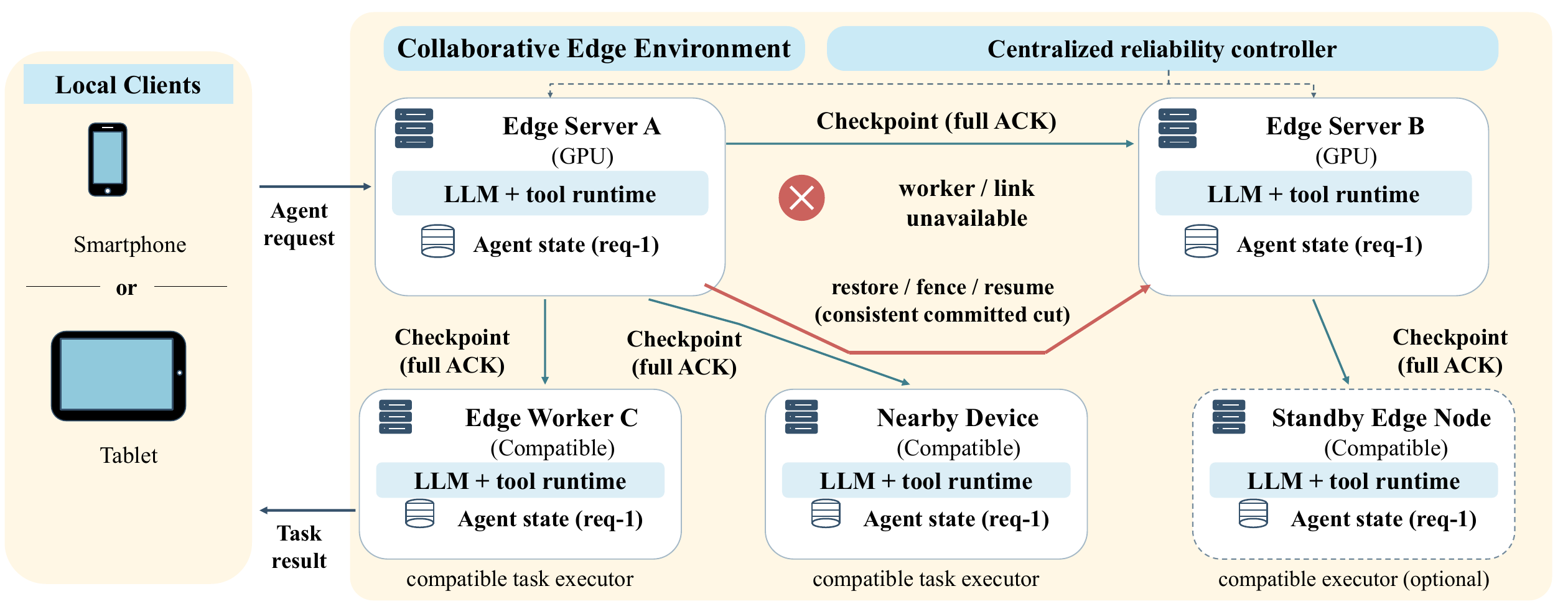}
    \caption{Collaborative edge architecture for AI agent execution.}
    \label{fig:agent_system}
\end{figure*}

Fig.~\ref{fig:agent_system} shows heterogeneous nodes $\mathcal{N}$ connected by links $\mathcal{E}$ on user-controlled infrastructure. Workers provide inference, isolated tool execution, and request-state storage. A logically centralized reliability controller manages client requests using observed workflows, node/link status, resource occupancy, and acknowledged checkpoints.

For subtask $v$ of request $i$, $\mathcal{C}_{iv}(t)\subseteq\mathcal{N}$ contains targets supporting its model, tools, state representation, runtime versions, and task-isolation permissions. Its authoritative executor is $p_{iv}(t)$, or $\bot$ when unassigned or suspended.

Models and tools are pre-provisioned. The control plane remains available during worker and data-link disruptions, retaining metadata and committed records while tracking referenced objects separately. Decisions use possibly delayed observations; control and history-access delays count toward execution or recovery time.

Reasoning and tool interactions reveal subsequent work and produce its required inputs. The revealed workflow of request $i$ is a dynamic DAG $\mathcal{A}_i(t)=(\mathcal{V}_i(t),\mathcal{E}_i(t))$, whose vertices are reasoning or tool subtasks and whose edges encode precedence and data dependencies. Outcomes change the remaining workload $W_i(t)$; repeated operations create new instances, preserving acyclicity. A subtask executes only when its required predecessor results are committed and accessible.

Let $n_i(t)$ count events committed at runtime-defined consistency boundaries. Committed records identify workflow versions, reasoning progress, tool outcomes, and referenced artifacts and effects. Later work depends on this history, but referenced objects may become unavailable after a disruption. Committed progress therefore differs from the state accessible for recovery.

Request $i$ arrives at $a_i$ with context $C_i^{\mathrm{ctx}}$, model/tool requirements $\Omega_i$, initial workload $W_i(a_i)$, and an external stopping rule. Its deadline is $D_i=a_i+d_i$, where $d_i$ is the target end-to-end latency, including queueing, inference, tools, communication, and contention. Policies share completion, termination, and reasoning/tool-use limits and observe only revealed work.

Execution requires both available compute and accessible or consistently reconstructible state. Protecting progress improves recovery options while consuming resources needed by ongoing work. A newer remote cut may save recomputation but require more transfer than an older local cut. These tradeoffs couple execution resources, state protection, and recovery choices under the remaining deadline budget.

\subsection{State Protection and Feasible Recovery}\label{subsec:sys_protection}

\begin{figure*}[t]
    \centering
    \includegraphics[width=\textwidth]{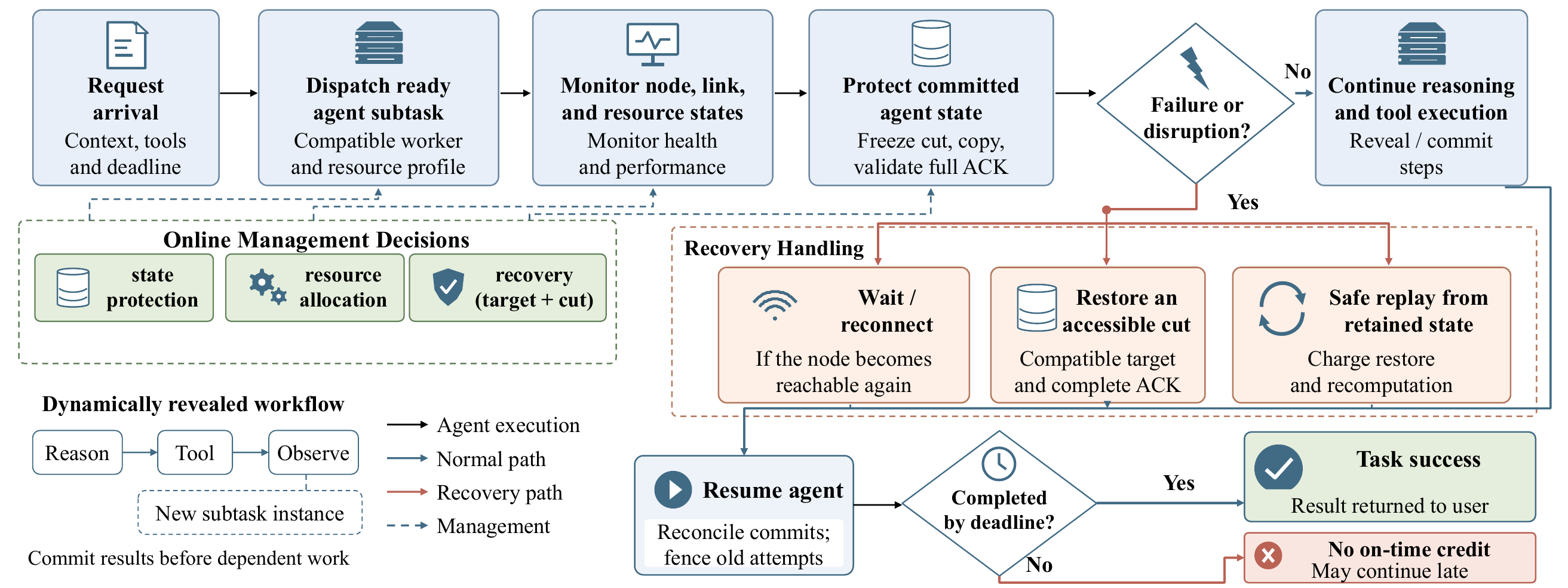}
    \caption{Agent execution, state protection, and recovery.}
    \label{fig:agent_recovery}
\end{figure*}

As shown in Fig.~\ref{fig:agent_recovery}, a checkpoint preserves a consistent cut of reasoning context, workflow status, tool results, files, and environment/version information. It excludes pre-provisioned weights and may include exportable KV state; local handles and external effects require supported reconstruction or reconciliation.

Let $\mathcal{O}_i(k)$ be the manifest of distinct versioned objects needed to restore request $i$ at committed cut $k$. Its logical payload in bytes is
\begin{equation}
    V_i(k)=\sum_{q\in\mathcal{O}_i(k)}\operatorname{size}(q).
    \label{eq:sys_state_payload}
\end{equation}
Here $\operatorname{size}(q)$ is the size of object $q$ in bytes. Versioned identities prevent incompatible states from being combined. Updates make $V_i(k)$ history-dependent and potentially nonmonotonic.

Let $k_{in}^{\mathrm{bk}}(t)$ be the largest complete compatible acknowledged cut retained at node $n$. With no checkpoint, zero is a bookkeeping sentinel and does not certify feasible initial-context reconstruction. Its committed-event lag is
\begin{equation}
    \begin{aligned}
    \delta_{in}(t)&=n_i(t)-k_{in}^{\mathrm{bk}}(t),\\
    0&\leq k_{in}^{\mathrm{bk}}(t)\leq n_i(t).
    \end{aligned}
    \label{eq:sys_freshness}
\end{equation}
Eq.~\eqref{eq:sys_freshness} measures committed-event lag, not missing bytes or recovery time. The catalog tracks complete versions, object locations, storage tiers, in-flight operations, and reserved capacity, including older complete cuts.

A protection operation freezes target cut $\bar{k}\leq n_i(t)$ at its start. If $\mathcal{B}_{in}(t)$ contains validated compatible destination objects, its incremental logical payload is
\begin{equation}
    \Delta V_{in}=\sum_{q\in\mathcal{O}_i(\bar{k})\setminus\mathcal{B}_{in}(t)}
    \operatorname{size}(q).
    \label{eq:sys_refresh_payload}
\end{equation}
A checkpoint becomes usable only after the complete manifest and objects are validated and acknowledged. Until then, partial bytes consume resources and the previous acknowledged cut remains valid. Source objects of the frozen cut are retained through transfer while later events may commit. A declared per-request replication cap bounds retained copies.

Failures change state availability in two ways: crashes can destroy volatile objects, while communication disruptions can leave intact objects temporarily inaccessible. Recovery at compatible target $n$ uses $\mathcal{K}_{in}^{\mathrm{avail}}(t)$, the set of complete compatible cuts available locally or through a feasible transfer. The selected cut satisfies
\begin{equation}
    \kappa_i(t)\in\mathcal{K}_{in}^{\mathrm{avail}}(t)\cup\{0\},
    \label{eq:sys_recovery_cut}
\end{equation}
subject to resource, artifact-availability, version, and replay conditions. Remote cuts incur transfer and preparation costs; locally accessible cuts support recovery independently of the primary link. Initial reconstruction, $\kappa_i(t)=0$, requires retained inputs and effects that permit safe replay.

Recovery restores the selected consistent cut, reconciles durable results, and fences obsolete attempts before execution resumes. Replay or recomputation from older cuts must respect committed tool outcomes and artifact versions, even when reasoning outputs change. External effects require operation identities and supported idempotency, outcome queries, or compensation; indeterminate irreversible effects must be resolved before replay. Recovery time includes detection, controller delay, queueing, transfer, restoration, replay, reconciliation, and resumption, accounting for overlap and further disruptions.

\subsection{Joint Management and Resource Constraints}\label{subsec:sys_network}\label{subsec:sys_management}

At each event-driven epoch, the controller jointly selects deployment-supported execution actions, compute/bandwidth reservations, checkpoint creation, refresh, retention or removal, and recovery targets and cuts.

Normal execution, state protection, rerouting, and recovery generate the flow set $\mathcal{F}(t)$. For each link or shared contention domain $e$, let $B_e(t)$ denote its usable capacity in bytes per second. Each flow $f\in\mathcal{F}(t)$ follows path $\mathcal{P}_f$ and receives effective goodput $u_f(t)$, also in bytes per second, subject to
\begin{equation}
    \sum_{f:e\in\mathcal{P}_f}u_f(t)\leq B_e(t),
    \qquad u_f(t)\geq0.
    \label{eq:sys_bandwidth}
\end{equation}
The constraint limits the aggregate goodput of flows sharing $e$, accounting for contention between execution and state transfers. An unavailable required endpoint or path provides no transfer progress. Failure and connectivity may vary over time and be correlated; risk estimates cover a declared horizon without revealing future faults.

At node $n$, compute pool $r$ has available capacity $G_n^r(t)$. Let $\gamma_{ivn}^r(t)$ be the allocation to subtask $v$ of request $i$, and $\gamma_n^{r,\mathrm{mgmt}}(t)$ the aggregate protection and recovery demand in the same pool. The binary variable $x_{ivn}(t)$ equals one when $n$ is the authoritative active executor of subtask $v$ of request $i$, and zero otherwise. CPU and accelerator pools are separate when applicable. Feasible allocations satisfy
\begin{equation}
    \begin{aligned}
    \sum_{i,v}\gamma_{ivn}^r(t)+\gamma_n^{r,\mathrm{mgmt}}(t)&\leq G_n^r(t),\\
    0\leq\gamma_{ivn}^r(t)&\leq x_{ivn}(t)G_n^r(t),
    \end{aligned}
    \label{eq:sys_compute}
\end{equation}
The first inequality charges task execution and protection/recovery against the same compute budget. The second permits task allocation only on its authoritative active executor. Protection and recovery demand remains included when a subtask is suspended. Profiled service rates depend on task type, allocation, and contention; linear scaling with allocation is not assumed.

For storage tier $h$ at node $n$, let $M_{nh}$ denote capacity and $W_{nh}$ the space occupied by resident model weights. Let $\mathcal{Q}_{nh}(t)$ be the set of distinct allocated state objects, with $\operatorname{size}_h(q)$ denoting the storage occupied by object $q$ in tier $h$. The term $M_{nh}^{\mathrm{other}}(t)$ covers workspace, staging, and background occupancy. Storage feasibility requires
\begin{equation}
    W_{nh}+\sum_{q\in\mathcal{Q}_{nh}(t)}\operatorname{size}_h(q)
    +M_{nh}^{\mathrm{other}}(t)\leq M_{nh}.
    \label{eq:sys_storage}
\end{equation}
The constraint counts each physical allocation once, including primary/backup objects, overlapping source/destination residency, and partial transfers. GPU memory, host memory, and workspace storage are separate tiers when used. Residual budgets deduct existing occupancy. Exogenous state or capacity loss can suspend or fail work; it does not create free resources or preserve destroyed objects.

Assignment and compatibility require

\begin{equation}
    \begin{aligned}
    \sum_{n\in\mathcal{N}}x_{ivn}(t)&\leq1,\\
    x_{ivn}(t)&\leq\mathbf{1}\{n\in\mathcal{C}_{iv}(t)\}.
    \end{aligned}
    \label{eq:sys_assignment}
\end{equation}

Eq.~\eqref{eq:sys_assignment} prevents multiple or incompatible authoritative executors. New roles also require committed accessible predecessors, consistent state, and actual resource reservation. Rerouting changes $p_{iv}(t)$ only after destination readiness is acknowledged and old authority fenced; recovery likewise requires successful reconstruction. Duplicate attempts cannot both commit authoritative subtask results.

An action initiates work rather than completing it instantaneously. For remaining transfer bytes $w_f(t)$,
\begin{equation}
    w_f(t+\Delta)=\left[w_f(t)-\int_t^{t+\Delta}u_f(\tau)\,d\tau\right]_+.
    \label{eq:sys_transfer_progress}
\end{equation}
Here $[z]_+=\max\{0,z\}$. Eqs.~\eqref{eq:sys_bandwidth} and \eqref{eq:sys_transfer_progress} charge transfer service under shared capacity. Acknowledged copies survive only as allowed by their storage tier and fault type. Interrupted operations retain occupied resources until completion or explicit release.

Let $X_t$ collect revealed workflows, execution locations, committed progress, checkpoint/object availability, queues, in-flight work, resource occupancy, and node/link state. For management action $U_t$ and disturbances or newly observed Agent outcomes $\Xi_t$,
\begin{equation}
    X_{t+\Delta}=F(X_t,U_t,\Xi_t).
    \label{eq:sys_transition}
\end{equation}
The transition $F$ represents the common physical execution/accounting rules. Actions alter queues, contention, and recoverability; a changed backup placement does not change an exogenous failure realization.

\subsection{Problem Formulation}\label{subsec:sys_objective}

Consider an externally generated cohort $\mathcal{R}$ of $N>0$ submitted requests, each visible only upon arrival. Every policy uses the same cohort and stopping rules. Rejected, failed, cancelled, and late requests remain in the denominator. Let $C_i^\pi$ be the absolute client-delivery time of a complete protocol-consistent output under policy $\pi$, or $+\infty$ if none is delivered. Consistency covers workflow continuation, checkpoints, and safe replay. Define $Y_i^\pi=\mathbf{1}\{C_i^\pi\leq D_i\}$ and $Z^\pi=N^{-1}\sum_{i\in\mathcal{R}}Y_i^\pi$.

A soft deadline determines on-time success. A late request may continue consuming compute, storage, and bandwidth until completion, observed cancellation, or explicit termination invokes the cleanup rules. Observed cancellation invokes the user stopping rule rather than failure-driven recovery. Thus the scoring window and the real resource lifetime are distinct.

Observation covers each deadline or an earlier terminal outcome, determining every on-time result.

Let $H_t$ be the available observation/action history and $\Pi_{\mathrm{adm}}$ the nonanticipative policies accepted by the resource-feasible executor. We seek an admissible online policy that jointly selects execution resources, state-protection operations, and feasible recovery targets and cuts to maximize the expected task success rate:

\begin{equation}
    \underset{\pi\in\Pi_{\mathrm{adm}}}{\operatorname{maximize}}
    \quad J(\pi)=\mathbb{E}[Z^\pi]=\frac{1}{N}\sum_{i\in\mathcal{R}}\Pr(C_i^\pi\leq D_i),
    \label{eq:sys_objective}
\end{equation}

subject to Eqs.~\eqref{eq:sys_bandwidth}--\eqref{eq:sys_storage} and \eqref{eq:sys_assignment}, together with the dependency, checkpoint-consistency, accessible-cut, safe-replay, and authority-transition conditions. The policy selects $U_t\sim\pi(\cdot\mid H_t)$ and execution follows Eq.~\eqref{eq:sys_transition}.

The expectation covers request characteristics, reasoning/tool outcomes, failures, network variation, and policy randomization. The equality in Eq.~\eqref{eq:sys_objective} holds for a fixed-size cohort without requiring independent request outcomes. State availability and completion time are evaluated jointly, not by multiplying marginal probabilities. Recovery interruption, latency, transferred bytes, and storage overhead are complementary metrics, not arbitrary penalties replacing Eq.~\eqref{eq:sys_objective}. Section~\ref{sec:solution} develops a world model-based MPC method for this problem.

\section{World Model-based Model Predictive Control}\label{sec:solution}

We develop world model-based model predictive control (WorldMPC) for reliable agent execution by jointly modeling workflow progress and recoverable state. The model generates execution and recovery trajectories that respect committed history. These trajectories guide joint resource, protection, and recovery decisions for the task success objective in Section~\ref{system-model-and-problem-formulation}.

\subsection{Partially Observable Semi-Markov Decision Process}\label{subsec:meth_decision_process}

We model online execution management as a partially observable semi-Markov decision process (POSMDP)~\cite{White1976POSMDP,Zhang2017POSMDP}, with partially observed execution states and variable intervals between decisions.

\emph{Latent state and decision epochs.}
At decision epoch $k$, the latent state is $S_k=(t_k,X_{t_k},\eta_k)$, where $t_k$ is physical time, $X_{t_k}$ is the execution state, and $\eta_k$ captures additional information needed for the Markov property, such as service ages, failure modes, and pending notifications. The state also tracks on-time delivery status. We assume the augmented state is Markov under management actions.

Decisions are triggered by request arrivals, subtask completions, checkpoint acknowledgments, detected disruptions, or periodic control timers. The interval $\Delta t_k=t_{k+1}-t_k$ varies across epochs. The semi-Markov transition kernel is
\begin{equation}
\begin{aligned}
Q(ds',d\delta\mid s,u)
=\Pr\{&S_{k+1}\in ds',\\
&\Delta t_k\in d\delta\mid S_k=s,U_k=u\}.
\end{aligned}
\label{eq:meth_posmdp_transition}
\end{equation}
Here, $s$ and $s'$ denote current and next latent states, $u$ the action, and $\delta$ elapsed time.

\emph{Actions and objective.}
To connect the POSMDP reward with the task success rate objective, consider a fixed evaluation cohort $\mathcal{R}$ of $N$ requests. Let $\ell_i(s)\in\{0,1\}$ indicate whether request $i$ has achieved valid on-time delivery in state $s$. The indicator is nondecreasing. The stage reward is
\begin{equation}
r(s,u,s',\delta)=\frac{1}{N}\sum_{i\in\mathcal{R}}[\ell_i(s')-\ell_i(s)].
\label{eq:meth_posmdp_reward}
\end{equation}
Summing these rewards over transitions from the initial state until all on-time outcomes are resolved gives the task success rate. The evaluation cohort is fixed, but future requests remain unknown to the online controller.

\emph{Online Decision Rule.}
Let $H_{t_k}$ contain observed telemetry, events, committed execution records, and previous actions. The controller uses this history to select a feasible action without computing the exact POSMDP belief:
\begin{equation}
\begin{aligned}
\sigma_k &= \operatorname{Enc}(H_{t_k}),\\
U_k &= \mu_{\phi}(\sigma_k)
\in\mathcal{U}_{\mathrm{feas}}(H_{t_k}),
\end{aligned}
\label{eq:meth_policy}
\end{equation}
Here, $\operatorname{Enc}$ constructs the observed state $\sigma_k$, and $\mathcal{U}_{\mathrm{feas}}$ is the feasible action set. WorldMPC implements $\mu_{\phi}$ using world-model parameters $\phi$, without a separately trained policy. The notation $U_k$ abbreviates $U_{t_k}$ in the system model.

\subsection{World Model}\label{subsec:meth_world_model}

\begin{figure}[t]
    \centering
    \includegraphics[width=\columnwidth]{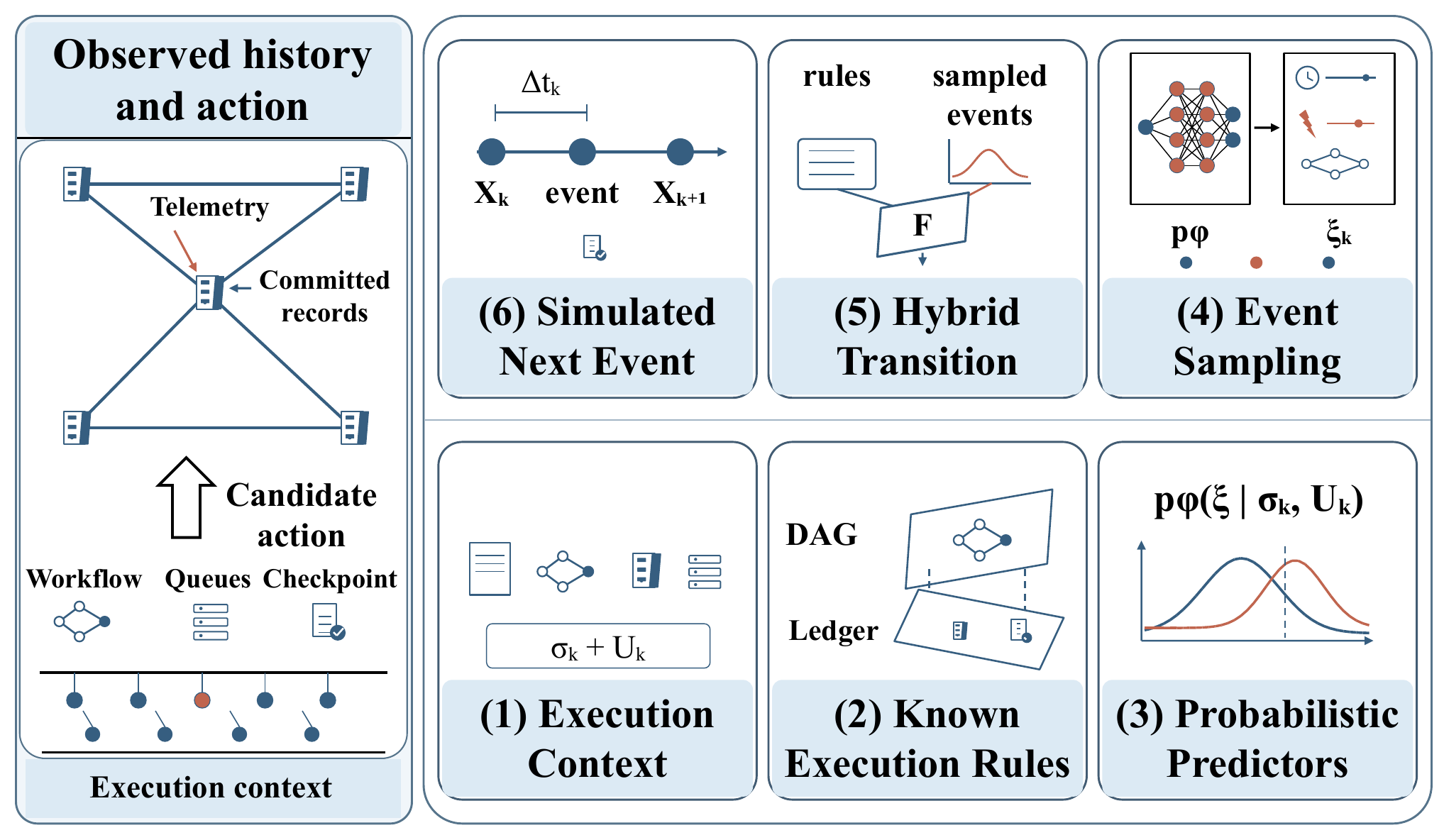}
    \caption{Hybrid world model for AI agent execution.}
    \label{fig:meth_world_model}
\end{figure}

Our hybrid world model captures the coupled evolution of workflow progress, recoverable state, and shared resources. Execution produces new state, protection consumes resources to preserve it, and disruptions change which states remain available for recovery. We model these effects using known execution rules and learned probabilistic dynamics, building on discrete-event system modeling~\cite{Cassandras2008DES} and probabilistic dynamics modeling~\cite{Chua2018PETS}. Fig.~\ref{fig:meth_world_model} illustrates how observed history and a candidate action define the execution context (1). Known rules govern workflow dependencies, resource accounting, and committed effects (2), while probabilistic predictors model execution times, disruptions, and workflow evolution (3). Event variables are sampled from the predictors (4) and processed under known rules by the hybrid transition (5), producing the next simulated event, elapsed time, execution progress, and recoverable state (6).

\textbf{Known Execution Rules.} The rule component implements the transition and accounting mechanism $F$ in Section~\ref{subsec:sys_management}. It updates task dependencies, role ownership, resource reservations, transfer progress, and acknowledged checkpoints. A usable recovery cut requires the complete manifest and objects, consistent committed effects, and a feasible restoration or replay path.

\textbf{Probabilistic Transition Model.} Let $\xi_k$ collect uncertain quantities needed to advance execution, including remaining service times, node and link changes, recovery costs, and newly revealed workflow information. The learned component samples

\begin{equation}
    \begin{aligned}
    \widehat{\xi}_k\sim
    &\widehat{p}_{\phi}(\cdot\mid\sigma_k,U_k),\\
    (\widehat{X}_{k+1},\widehat{\Delta t}_k)
    &=\mathcal{F}(\widehat{X}_k,U_k,\widehat{\xi}_k),
    \end{aligned}
    \label{eq:meth_hybrid_transition}
\end{equation}

where $\mathcal{F}$ is the event-driven implementation of $F$. Here, $\widehat{X}_k$ and $\widehat{X}_{k+1}$ are simulated execution states, $\widehat{\Delta t}_k$ is the predicted time to the next event, and $\widehat{\xi}_k$ contains sampled uncertain event variables. It advances execution, transfers, and resource occupancy to the next event, with the control timer bounding the interval between decisions. Observations and acknowledged records initialize simulation.

Predictor features describe subtask and tool types, model and device identifiers, resource allocations, elapsed service, queues, checkpoint payload and age, and recent network and failure observations. Conditioning on elapsed service supports residual-duration prediction for running operations. Shared history or scenario variables can represent correlated disruptions. Workflow predictions describe supported dependency extensions and workload or state-size changes; actual tool outputs and committed files enter the system through execution.

\textbf{Model Training.} Each record $\zeta_j=(\sigma_j,U_j)$ uses the history available before an observed event target $y_j$. Targets include event times, types, and observed workflow, resource, or recovery changes. The likelihood uses the same event model as future sampling, with masks and marginalization for unavailable fields. Training and validation splits keep requests with shared congestion and disruptions together.

For fully observed event targets and administratively censored event intervals, respectively, the training objective is

\begin{equation}
    \begin{split}
    \mathcal{L}(\phi)=-\frac{1}{|\mathcal{D}|}\bigg[&
    \sum_{j\in\mathcal{D}_{\mathrm{obs}}}
    \log\widehat{p}_{\phi}(y_j\mid\zeta_j)\\
    &+\sum_{j\in\mathcal{D}_{\mathrm{cens}}}
    \log\widehat{S}_{\phi}(c_j\mid\zeta_j)\bigg],
    \end{split}
    \label{eq:meth_model_loss}
\end{equation}

Here, $\mathcal{D}_{\mathrm{obs}}$ and $\mathcal{D}_{\mathrm{cens}}$ contain observed and censored event records, respectively, and $\mathcal{D}$ is the complete training set. For each censored record $j$, $c_j$ is the observed event-free interval, and $\widehat{S}_{\phi}(c_j\mid\zeta_j)$ is the predicted probability that no event occurs within this interval. Completion and disruption are distinct event types. A crash supervises the disruption target, while an incomplete record contributes its observed fields or justified censoring term. Each target contributes once. For each training mini-batch, the parameters are updated by

\begin{equation}
    \phi\leftarrow\phi-\eta_{\phi}\nabla_{\phi}\mathcal{L}(\phi),
    \label{eq:meth_model_update}
\end{equation}

where $\eta_{\phi}$ is the learning rate. We select the model using a declared validation criterion for event-time likelihood and disruption calibration across workload and failure conditions. The selected model stays fixed within each planning call; trace-based updates take effect between calls.

In the evaluated implementation, duration distributions and event probabilities or rates are fitted to observations, with declared priors for unmeasured processes. Section~\ref{sec:performance-evaluation} specifies these components.

Algorithm~\ref{algo:wm_training} summarizes world-model training. Lines 1--3 construct history-conditioned records, separate training and validation runs, and initialize the predictors. Lines 4--8 fit the predictors using observed and censored event targets in Eq.~\eqref{eq:meth_model_loss}. Lines 9--12 select and freeze the planning model using predictive likelihood and disruption calibration.

\begin{algorithm}[t]
    \caption{Supervised Training of the Hybrid World Model}
    \label{algo:wm_training}
    \begin{algorithmic}[1]
        \Require Execution traces $\mathcal{D}$, training epochs $E$, learning rate $\eta_{\phi}$
        \Ensure Trained model $\widehat{p}_{\phi}$ and fixed feature definitions
        \State Build history-conditioned records and observed-target masks\label{line:wm_records}
        \State Separate execution runs into training and validation sets
        \State Initialize probabilistic predictor parameters $\phi$
        \For{epoch $=1$ to $E$}
            \For{each training mini-batch}
                \State Evaluate the observed and censoring likelihoods in Eq.~\eqref{eq:meth_model_loss}
                \State Update $\phi\leftarrow\phi-\eta_{\phi}\nabla_{\phi}\mathcal{L}(\phi)$\label{line:wm_update}
            \EndFor
            \State Evaluate predictive likelihood and disruption calibration on validation runs
            \State Retain the model version with the best declared validation criterion
        \EndFor
        \State Freeze the selected model version for online planning\label{line:wm_freeze}
        \State \Return $\widehat{p}_{\phi}$ and the feature definitions
    \end{algorithmic}
\end{algorithm}

\subsection{State and Action Spaces}\label{subsec:meth_state}\label{subsec:meth_action}

We define the controller's observed state representation at epoch $k$ as
\begin{equation}
\sigma_k=(g_k,d_k,e_k,q_k,c_k),
\label{eq:meth_state}
\end{equation}

Here, $g_k$ represents revealed workflows and execution progress, $d_k$ tracks remaining deadlines, $e_k$ describes observed device and network conditions, $q_k$ records execution and transfer queues, and $c_k$ represents checkpoint availability and recovery states.

At each decision epoch, the controller selects an action or a jointly feasible batch of actions. The action is defined as

\begin{equation}
    U_k=(u_k,\vartheta_k,z_k,\rho_k),
    \label{eq:meth_action}
\end{equation}

where $u_k$ denotes the action type, $\vartheta_k$ the request, subtask, and target, $z_k$ checkpoint operations and destinations, and $\rho_k$ compute/bandwidth reservations.

The execution action types and parameter domains are defined by the deployment scenario. At each decision epoch, the observed workflow state, resource availability, and recovery conditions determine which actions are feasible. Concurrent actions must satisfy aggregate resource, compatibility, and recovery constraints. When no feasible action is available, the executor waits for the next event or control timer while accepted work continues.

\subsection{Reward Estimation}\label{subsec:meth_success}

The planner estimates the expected on-time delivery reward for currently observed requests. Let $\mathcal{I}_k$ be the set of requests whose on-time outcomes remain unresolved at epoch $k$. For a candidate feedback plan $\Pi$, let $\widehat{C}_{i,m}^{\Pi}$ denote the valid client-delivery time of request $i$ in world-model rollout $m$, or $+\infty$ if valid delivery does not occur. Using $M$ complete rollouts, the estimated reward is

\begin{equation}
    \widehat{J}_k(\Pi)=
    \frac{1}{M|\mathcal{I}_k|}
    \sum_{m=1}^{M}\sum_{i\in\mathcal{I}_k}
    \mathbf{1}\{\widehat{C}_{i,m}^{\Pi}\leq D_i\}.
    \label{eq:meth_score}
\end{equation}

Valid delivery requires workflow consistency and safe recovery. All candidate plans use the same request set $\mathcal{I}_k$, making the estimated reward comparable across plans. This score is a conditional planning estimate, while the reported task success rate covers all submitted requests. Ties are resolved by predicted recovery interruption, transferred bytes, and fixed candidate order. New arrivals trigger replanning; an empty $\mathcal{I}_k$ invokes the feasible continuation policy.

\subsection{World Model-based Model Predictive Control}\label{subsec:meth_mpc}

\begin{figure*}[t]
    \centering
    \includegraphics[width=\textwidth]{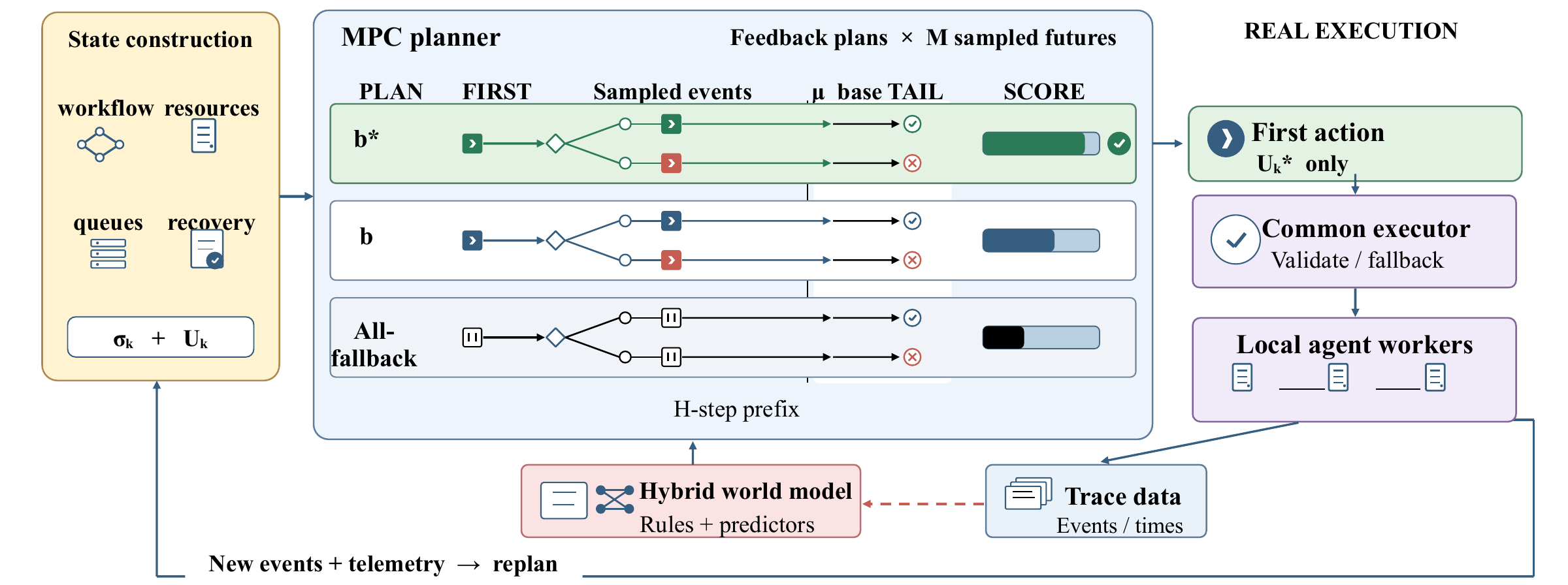}
    \caption{WorldMPC framework.}
    \label{fig:meth_framework}
\end{figure*}

Fig.~\ref{fig:meth_framework} shows how WorldMPC combines model training, predictive planning, and execution feedback to jointly select resource allocation, state protection, and recovery actions. Historical traces train the world model's predictors, while online observations form the controller state. The planner evaluates feedback plans and the all-fallback plan over $M$ sampled futures. Feedback plans use an $H$-step optimized prefix and, when needed, a common continuation policy, with scores based on valid on-time delivery.

\begin{algorithm}[t]
    \caption{World Model-based MPC for Reliable Agent Execution}
    \label{algo:wm_mpc}
    \raggedright
    \begin{algorithmic}[1]
        \Require $\widehat{p}_{\phi}$, $\mu_{\mathrm{base}}$, caps $B,K,H,M,L$, budget $\Omega_k$
        \Ensure Feasible execution, protection, and recovery
        \While{the service is active}
            \State Observe events, telemetry, ACKs, and cancellations\label{line:mpc_observe}
            \State Construct $\sigma_k$ and fixed request set $\mathcal{I}_k$
            \State Select the all-fallback plan
            \If{$\mathcal{I}_k\neq\emptyset$, actions feasible, context supported}
                \State Start timer; set $\mathcal{P}_k=\emptyset$
                \State Generate $\leq B$ plans using $\mathcal{L}_K$ and $H$\label{line:mpc_candidates}
                \For{each plan $\Pi^b$ while budget remains}
                    \State Initialize rollout outcomes
                    \For{$m=1$ to $M$ while budget remains}
                        \State Initialize state/ledger from observations
                        \State Advance by estimated planning delay
                        \State Run $\Pi^b$ until resolved or $H$ events\label{line:mpc_prefix}
                        \If{scoring outcomes unresolved}
                            \State Run $\mu_{\mathrm{base}}$ for $\leq L$ events\label{line:mpc_tail}
                        \EndIf
                        \If{rollout incomplete or unsupported}
                            \State Discard $\Pi^b$
                            \State \textbf{break} rollout loop
                        \Else
                            \State Record each valid delivery once
                        \EndIf
                    \EndFor
                    \If{all $M$ rollouts complete within budget}
                        \State Score by Eq.~\eqref{eq:meth_score}; add $\Pi^b$ to $\mathcal{P}_k$\label{line:mpc_score}
                    \EndIf
                \EndFor
                \If{$\mathcal{P}_k\neq\emptyset$}
                    \State Select highest score with declared tie breakers
                \EndIf
            \EndIf
            \State Refresh observations; compile first action only\label{line:mpc_first}
            \State Execute valid action or feasible fallback;
            \Statex \hspace{\algorithmicindent} otherwise retain ongoing execution
            \State Log planning time, actions, events, and outcomes
            \State Await event or timer while accepted work continues\label{line:mpc_replan}
        \EndWhile
    \end{algorithmic}
\end{algorithm}

\textbf{Feasible Candidate Generation.} The controller builds $\mathcal{L}_K(\sigma)$ from deployment-supported action types and parameter domains, combining eligible subtasks, compatible targets, retained recovery cuts, resource profiles, and checkpoint operations. Protection and resource adjustments may remain feasible during execution. Tuples are ordered by request deadline, request identifier, target identifier, resource-profile identifier, and protection choice. Within the remaining time budget, the generator lazily inspects at most $K$ tuples and retains feasible actions, with any feasible fallback retained independently. This ordering governs enumeration, not the task success objective.

Except for the all-fallback plan, each candidate is a rank sequence $\boldsymbol{r}^{b}=(r_0^b,\ldots,r_{H-1}^b)$, with $r_h^b\in\{1,\ldots,K\}$. At each of the $H$ simulated decision steps, it rebuilds the option list from the trajectory history and applies the requested rank or falls back if that rank is unavailable. The plan thus responds to revealed completions and failures. Generation includes the all-fallback plan, covers distinct feasible first actions as the cap permits, and uses seeded rank-sequence sampling for remaining candidates. At most $B$ generation attempts are made, and duplicates are removed.

\textbf{Monte Carlo Rollouts.} Each candidate is evaluated in $M$ joint execution scenarios sampled from Eq.~\eqref{eq:meth_hybrid_transition}. Each trajectory follows task execution, state protection, and possible recovery while maintaining consistency with committed reasoning outputs, tool results, and files. It accounts for shared-resource contention and the transfer, restoration, and replay required along a feasible recovery path. Every simulated decision is revalidated. Common random seeds are used for candidate evaluation, although different actions may produce different event sequences.

After the optimized $H$-step prefix, each trajectory uses the same continuation policy $\mu_{\mathrm{base}}$ to resolve complete request delivery. A complete score is available once every request in $\mathcal{I}_k$ has delivered or missed its on-time target, by $T_k^{\mathrm{score}}=\max_{i\in\mathcal{I}_k}D_i$ at the latest. A per-request ledger records each delivery once.

The continuation policy preserves running work and acknowledged checkpoints. At supported scheduling boundaries, interrupted and ready requests follow earliest deadline, with request age and identifier breaking ties. For interrupted work, it chooses a compatible target and cut minimizing predicted restoration plus remaining execution; for ready work, it minimizes predicted feasible finish time over the same resource profiles. Infeasible work waits while other feasible work proceeds. Late ready work is served when feasible on-time work cannot use the resources, so sustained overload can postpone it.

\textbf{Bounded Planning and Runtime Safeguards.} Caps $B$, $M$, $H$, $L$ and budget $\Omega_k$ bound planning. Candidates qualify only if all $M$ scoring rollouts finish within these limits and supported contexts. Exclusion does not mark real requests as failed; if none qualifies, the controller uses $\mu_{\mathrm{base}}$. Late work retains resources, running tools remain non-preemptive, and observed cancellations follow the stopping rule. Completed-rollout, exclusion, and fallback counts track search limits and simulation costs.

Planning and state access consume request time while accepted work continues. Rollouts advance existing work under current allocations through the declared planning-delay estimate. Without a feasible action, they advance to the next event or timer. Before execution, the controller refreshes telemetry and validates the selected first action or uses a feasible fallback. If neither is available, accepted work continues until the next event or timer. Actual planning time is charged to the execution clock.

For bounded workflows and option lists, rollout evaluation costs $O(BM(H+L)c_{\mathrm{step}})$, excluding state construction and candidate generation. Each event step includes predictor evaluation, state updates, option-list reconstruction, and feasibility checks. Evaluated parametric predictors run on the controller CPU without additional LLM calls; sequential evaluation retains only active trajectories. Measured budgets include generation and scoring. Experiments report caps, resource profiles, the validation criterion, delay estimates, and fallback frequency.

Algorithm~\ref{algo:wm_mpc} implements the online decision loop. Lines 1--7 construct the observed state, initialize the fallback, and generate feasible plans. Lines 8--27 simulate the optimized prefix and common continuation, account for planning delay, and score only candidates with $M$ complete rollouts. Lines 28--31 select the highest-scoring plan or retain the fallback if none qualifies. Lines 32--36 refresh observations, validate and execute the first action or a feasible fallback, record outcomes and overhead, and wait for the next event or control timer.

\setlength{\textfloatsep}{10pt plus 2pt minus 2pt}
\setlength{\dbltextfloatsep}{10pt plus 2pt minus 2pt}
\setlength{\floatsep}{8pt plus 2pt minus 2pt}
\setlength{\dblfloatsep}{8pt plus 2pt minus 2pt}
\setlength{\abovecaptionskip}{6pt}

\section{Performance Evaluation}\label{sec:performance-evaluation}

We evaluate WorldMPC using Task Success Rate (TSR) in Section~\ref{subsec:sys_objective}: the fraction of all submitted requests whose complete, protocol-consistent outputs reach clients on time.

\subsection{Experimental Setting}\label{subsec:eval_setting}

\subsubsection{Dataset}

The public dataset is SWE-bench Verified, which contains $500$ issue tasks from $12$ repositories~\cite{jimenez2024swebench}. We construct simulation workflows by mapping issue-text length to workflow depth, nominal operation work, and execution-state size. The workflows alternate reasoning and tool stages and end in a receipt.

The synthetic dataset equally represents reasoning-only, read-only tool, and read-write tool workflows. Each request has three to five sequentially revealed stages, with one operation executed at a time; read-write requests end in a receipt. Operation work has multiplicative lognormal variability, and state size grows after completed operations. Arrivals follow spaced or burst patterns. Relative deadlines are $d_i=\alpha(1.5W_i+12)$ seconds, where $W_i$ is total nominal work and the deadline factor $\alpha$ scales the reference deadline $(1.5W_i+12)$ seconds.

\subsubsection{Simulation Environment}

The simulation has two heterogeneous workers sharing a $100$ Mbps link. Their fastest-to-slowest service-rate ratio is two, with total nominal compute normalized to two reference-worker units. Each worker provides separate reasoning and tool pools and $384$ MB of state memory. Execution, checkpoint preparation, restoration, and data transfers obey shared resource constraints enforced by a common executor. Running operations are non-preemptive. Feasible unfinished on-time work takes priority over late ready work, while late execution retains its resource occupancy.

WorldMPC selects compute fractions from $\{1,0.5\}$ and can select protection actions; the four baselines use fraction one and $8$-second periodic checkpointing at supported committed boundaries. Service rates scale with the allocated fraction raised to $0.8$. All methods share dependency checks, compatibility constraints, transfer accounting, and recovery rules. The comparison therefore evaluates the complete control policies, including their allocation and protection choices, rather than isolating predictive planning alone.

Five environments examine different sources of prediction difficulty: healthy operation, observable pressure, opaque failures, common outages, and misleading pressure. They vary worker failure risk, the informativeness of pressure signals, and correlated outages. Nonhealthy environments also include shared-link outages. These conditions are prescribed, not measured physical failure distributions.

The learned event predictors in the simulation use eight fitted scalars: a log-location and log-scale for each of the reasoning, tool, and receipt duration predictors, and two pressure-conditioned disruption rates. Duration predictors use censored lognormal likelihood with $100$ full-batch updates and learning rate $0.01$; disruption rates use exposure records and a weak Gamma prior. Training and validation contain $2400$ and $800$ synthetic observations, respectively. For $N$ training records and $E$ updates, fitting these fixed-size predictors costs $O(EN)$. The fitted model remains fixed during evaluation. Rollouts use worker-wise hazards and a prior for recovery of a currently down link.

Each dataset is evaluated across the five environments and two arrival patterns, using $27$ requests per run and $24$ paired random seeds. Methods share requests, worker profiles, and disruption calendars for each seed. Worker-count sweeps add capacity at fixed workload, whereas heterogeneity sweeps preserve total compute. The main parameter ranges appear in Table~\ref{tab:eval_parameters}. Simulations run in Python on an Intel Core i5-12600KF, with measured planning time charged to the execution clock. 

\begin{table}[t]
    \centering
    \caption{Principal simulation and planning parameters.}
    \label{tab:eval_parameters}
    \setlength{\tabcolsep}{5pt}
    \begin{tabular}{lc}
        \hline
        Parameter & Value \\
        \hline
        Requests per episode & $18,27,36$; transfer: $36,45$ \\
        Independent evaluation seeds & \EvalSeeds{}; $4$ \\
        Virtual workers & $2,4,8$; transfer: $4,6$ \\
        Shared bandwidth (Mbps) & $25,100,400$; transfer: $60$ \\
        Fastest/slowest service-rate ratio & $1,2,4$ \\
        State memory per worker (MB) & $384$ \\
Deadline factor $\alpha$ & $1.5,3,4.25,6$ \\

Failure-rate factor & $0,1,2$ \\

        Acknowledged remote replicas/request & At most $1$ \\
        Baseline protection interval (s) & $8$ \\
        Control-timer interval (s) & $2$ \\
        Candidate cap $B$ / option cap $K$ & $4/8$ \\
        Prefix depth $H$ / samples $M$ & $2/4$ \\
        Continuation-event cap $L$ & $300$ \\
        Planning budget (ms) & $20$ \\
        Per-request retry cap & $12$ attempts/operation \\
        Episode cleanup horizon (s) & $900$ \\
        \hline
    \end{tabular}
\end{table}

\subsection{Baselines}\label{subsec:eval_baselines}

We compare WorldMPC against four baselines. Heterogeneous Earliest Finish Time (HEFT) orders ready operations by upward ranks computed from computation and communication estimates on the currently revealed DAG. Earliest Deadline First (EDF) orders ready work by absolute request deadlines, with arrival order breaking ties. First In, First Out (FIFO) follows request arrival order. Failure Risk Heuristic (FRH) follows arrival order and favors placements with the lowest current risk estimated from observed pressure and failure history, breaking ties by estimated completion time.

\subsection{Experimental Results}\label{subsec:eval_results}

\subsubsection{World Model Training and Prediction}

The independent validation set measures event-time prediction and disruption calibration. Operation-time NLL is \EvalModelNLL{}, with mean absolute duration-ratio error \EvalModelDurationError{} on uncensored operations. Disruption prediction gives Brier score \EvalModelBrier{} and NLL \EvalModelDisruptionNLL{} over generated $10$--$35$-second exposure intervals. Censored records contribute survival likelihood terms. These values characterize the fitted parametric world model under the synthetic data distribution.

\subsubsection{Overall Performance}

\begin{figure}[!tb]
    \centering
    \includegraphics[width=\linewidth,trim=0bp 18.76bp 0bp 0bp,clip]{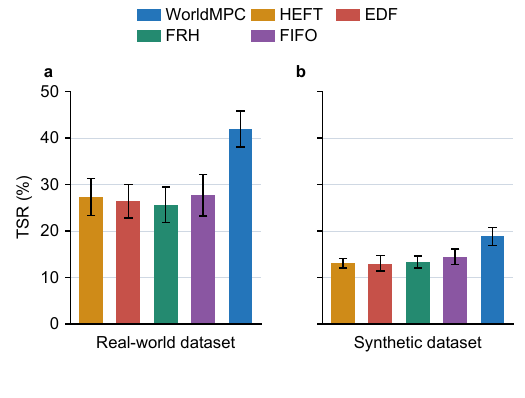}
    \caption{Task success rate on the real-world and synthetic datasets.}
    \label{fig:eval_overall_tsr}
\end{figure}

Fig.~\ref{fig:eval_overall_tsr}(a) compares TSR on the real-world dataset to evaluate WorldMPC's joint execution decisions. Compared with HEFT, EDF, FRH, and FIFO, WorldMPC increases TSR by approximately $54.1\%$, $57.9\%$, $63.3\%$, and $51.6\%$, respectively. The observed gains are consistent with considering execution time, resource contention, and recovery cost together when selecting actions.

Fig.~\ref{fig:eval_overall_tsr}(b) compares TSR on the synthetic dataset. Relative to HEFT, EDF, FRH, and FIFO, WorldMPC increases TSR by approximately $44.1\%$, $46.2\%$, $42.1\%$, and $30.4\%$, while reducing mean delivery latency by approximately $20.3\%$, $11.1\%$, $19.8\%$, and $15.8\%$, respectively. Both improvements are consistent with jointly accounting for waiting, execution, and recovery under shared resource constraints.

\begin{figure*}[!tb]
    \centering
    \includegraphics[width=\linewidth,trim=0bp 5.05bp 0bp 0.49bp,clip]{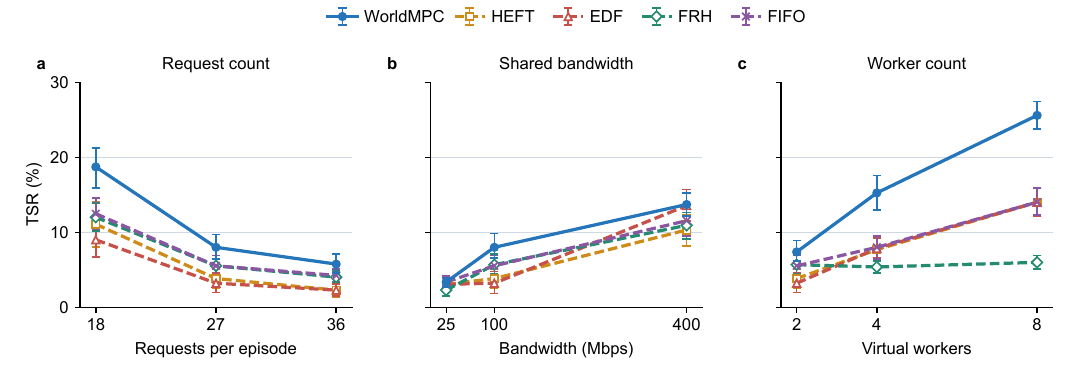}
    \caption{TSR versus the number of requests, shared bandwidth, and the number of workers.}
    \label{fig:eval_scaling}
\end{figure*}

\subsubsection{Effects of Workload and Resources}

Fig.~\ref{fig:eval_scaling}(a) evaluates how request count affects TSR under fixed resources. Increasing the request count from $18$ to $36$ reduces TSR as more execution and state-transfer operations compete for compute and bandwidth. At $36$ requests, WorldMPC achieves approximately $150.0\%$, $150.0\%$, $42.9\%$, and $35.1\%$ higher TSR than HEFT, EDF, FRH, and FIFO, respectively. WorldMPC evaluates candidate decisions using contention, recovery work, and remaining deadlines.

Fig.~\ref{fig:eval_scaling}(b) examines how shared bandwidth affects TSR when execution, protection, and recovery transfers compete for the same link. Higher bandwidth improves WorldMPC TSR. At $400$ Mbps, its TSR exceeds HEFT, EDF, FRH, and FIFO by approximately $32.8\%$, $1.1\%$, $25.4\%$, and $18.7\%$, respectively. The small observed gap to EDF suggests that deadline ordering becomes more competitive when communication pressure is reduced.

Fig.~\ref{fig:eval_scaling}(c) evaluates how WorldMPC uses additional workers under a fixed number of requests and shared bandwidth. Increasing the worker count improves WorldMPC TSR. At eight workers, WorldMPC achieves approximately $82.4\%$, $82.4\%$, $325.6\%$, and $82.4\%$ higher TSR than HEFT, EDF, FRH, and FIFO, respectively. Additional workers expand execution and recovery choices, which WorldMPC evaluates with resource allocation and remaining deadlines.

\begin{figure}[!tb]
    \centering
    \includegraphics[width=\linewidth,trim=0bp 1.54bp 0bp 0.52bp,clip]{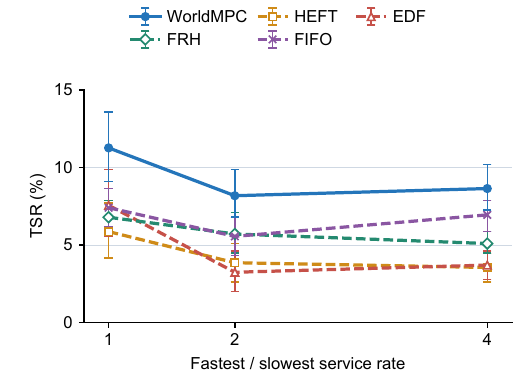}
    \caption{TSR under different levels of compute heterogeneity.}
    \label{fig:eval_heterogeneity}
\end{figure}

Fig.~\ref{fig:eval_heterogeneity} shows a nonmonotonic response to compute heterogeneity at fixed total capacity. At the largest speed ratio, WorldMPC achieves approximately $143.5\%$, $133.3\%$, $69.7\%$, and $24.4\%$ higher TSR than HEFT, EDF, FRH, and FIFO, respectively. A wider speed spread creates attractive fast-worker placements but can also concentrate contention on those workers. This trade-off is consistent with the nonmonotonic curve: greater heterogeneity alone does not ensure higher TSR, while coordinated decisions remain useful at the largest tested ratio.

\subsubsection{Reliability and Stateful Recovery}

\begin{figure}[!tb]
    \centering
    \includegraphics[width=\linewidth,trim=0bp 0bp 0bp 0bp,clip]{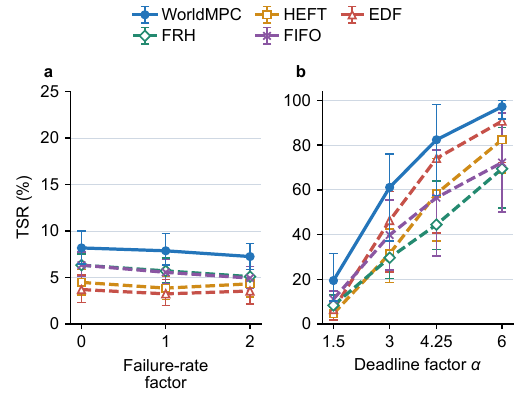}
    \caption{TSR under different failure-rate and deadline factors.}
    \label{fig:eval_reliability}
\end{figure}

 Fig.~\ref{fig:eval_reliability}(a) shows that stronger failures reduce on-time completion. Failures consume the remaining budget through waiting, transfer, restoration, and reconstruction. In the synthetic main comparison, WorldMPC reduces mean restoration interruption by approximately $3.9\%$, $6.6\%$, $11.9\%$, and $12.3\%$ relative to HEFT, EDF, FRH, and FIFO, respectively.

Because all methods share the recovery executor, the observed means reflect the recovery episodes produced by their scheduling and protection choices. On the real-world dataset, WorldMPC's mean interruption is approximately $6.8\%$ higher than HEFT and $6.8\%$ higher than FRH, despite its higher TSR. The resulting insight is that minimizing interruption duration alone is not equivalent to maximizing timely task completion; recovery must be evaluated together with the remaining execution work and deadline.

Fig.~\ref{fig:eval_reliability}(b) shows that TSR increases with the deadline factor $\alpha$. On average, WorldMPC achieves approximately $47.1\%$, $19.6\%$, $71.3\%$, and $44.8\%$ higher TSR than HEFT, EDF, FRH, and FIFO, respectively. WorldMPC jointly considers execution time, resource contention, and recovery costs when selecting actions within the remaining deadline budget.

\subsubsection{Generalization Analysis to Unseen Conditions}

The two combined configurations evaluate the same trained model and planning caps under changed workload and resource conditions. In the first configuration, WorldMPC matches HEFT and EDF, exceeds FRH by approximately $72.4\%$, and matches FIFO. In the second configuration, its TSR is approximately $25.0\%$, $25.0\%$, $78.6\%$, and $25.0\%$ higher than HEFT, EDF, FRH, and FIFO, respectively. WorldMPC remains competitive or better in these two configurations without retraining.

\subsubsection{Overhead and Component Analysis}

\begin{figure}[!tb]
    \centering
    \includegraphics[width=\linewidth,trim=0bp 13.37bp 0bp 7.02bp,clip]{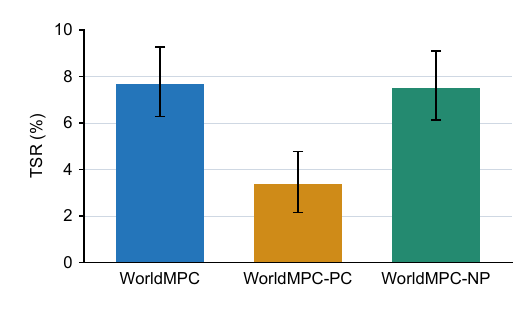}
    \caption{TSR with different state protection mechanisms.}
    \label{fig:eval_ablation}
\end{figure}

In the main synthetic comparison, WorldMPC averages \EvalPlanningMeanMS{} ms per controller callback, with fallback in \EvalFallbackPct{}\% of Monte Carlo planning calls. Callback time includes selections that launch no search; fallback uses actual search calls as its denominator. Cooperative budget checks occur between rollout events, so elapsed callback time can exceed $20$ ms. Checkpoint and total shared-link traffic average \EvalCheckpointMB{} and \EvalNetworkMB{} MB per submitted request, respectively.

Fig.~\ref{fig:eval_ablation} compares protection mechanisms while retaining sampled planning. Relative to WorldMPC-PC, full WorldMPC achieves approximately $125.8\%$ higher TSR with $39.3\%$ lower shared-link traffic. This combination is consistent with reducing periodic checkpoint transfers that compete with execution and recovery traffic. Relative to WorldMPC-NP, full WorldMPC achieves approximately $2.1\%$ higher TSR with $1.8\%$ higher traffic. WorldMPC-NP retains committed-log reconstruction, so removing additional checkpoints does not remove all recovery capability.

\subsubsection{Predictive Planning under Observable Failure Risk}

\begin{figure*}[!tb]
    \centering
    \includegraphics[width=\textwidth,trim=0bp 13.37bp 0bp 0bp,clip]{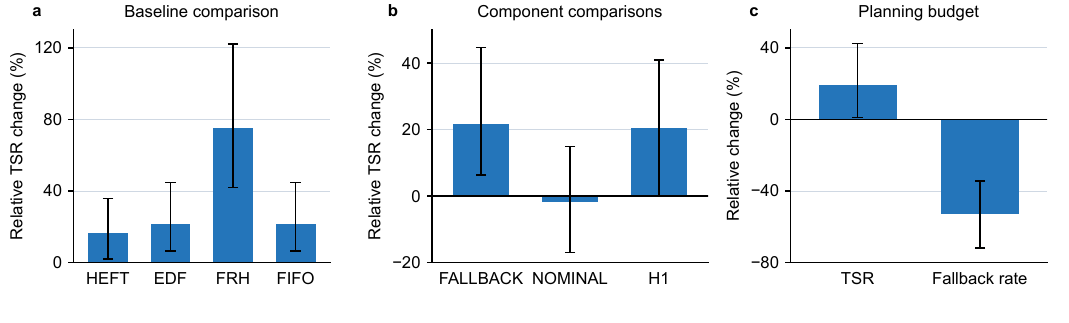}
    \caption{Predictive planning under observable failure risk.}
    \label{fig:predictable_resource_contention}
\end{figure*}

We further evaluate the contribution of predictive planning to TSR. We compare WorldMPC with the baselines and its variants. We also examine how the planning budget affects TSR and fallback. Planning and recovery costs are included in the evaluation.

Fig.~\ref{fig:predictable_resource_contention}(a) compares WorldMPC with HEFT, EDF, FRH, and FIFO. WorldMPC improves TSR by 16.7\%, 21.7\%, 75.0\%, and 21.7\%, respectively, relative to each baseline. WorldMPC evaluates how execution choices affect the completion of the remaining workflow. This allows it to balance execution speed against interruption and recovery costs when selecting actions to improve TSR.

Fig.~\ref{fig:predictable_resource_contention}(b) compares WorldMPC with fallback-only execution, fixed nominal predictors, and a one-command prefix. These variants remove trajectory search, replace the learned predictors with fixed priors, or shorten the optimized action prefix, respectively. WorldMPC improves TSR by 21.7\% over fallback-only execution and by 20.4\% over the one-command prefix. Evaluating alternative execution plans allows WorldMPC to consider how current actions affect subsequent execution and recovery.

Fig.~\ref{fig:predictable_resource_contention}(c) compares the two planning budgets. Relative to the shorter budget, the larger budget improves TSR by 19.1\% and reduces the fallback rate by 52.6\%. A larger budget permits full evaluation of more candidate plans, giving WorldMPC more opportunities to improve TSR.

\subsection{Real-world Testbed Evaluation}\label{subsec:eval_testbed_mechanisms}

\subsubsection{Testbed Setup}

We developed PolyLink, a collaborative edge platform for distributed LLM inference across heterogeneous devices~\cite{liu2025polylink}. It connects more than 100 consumer-grade GPUs from laboratories at dozens of universities~\cite{zhang2026efficient}. Such infrastructure can provide inference resources for AI agents, whose multi-step execution also requires progress and recoverable state to be maintained under disruptions. We implement WorldMPC in an AI agent execution prototype to coordinate resource allocation, state protection, and recovery in this setting.

A reliability controller dispatches Agent stages to isolated tool workspaces for file reading, reasoning, arithmetic or issue-metadata processing, and result validation. Committed results and validation outcomes reveal subsequent stages or repairs. We inject Agent-process termination and loss of worker-side checkpoint storage during execution. Checkpoint-based recovery verifies the saved cut before restoring the Agent context and files. Tool stages are replayed from the restored state, and the controller rejects results from obsolete attempts.

The real-world testbed uses two RTX 4090 worker hosts connected to existing model services. Each worker admits one Agent stage at a time. Tool containers use two profiles: $0.5$ CPU with $256$ MiB memory and $1$ CPU with $512$ MiB memory. Checkpoint storage is bounded at $64$ MiB per worker. Export and import share one serialized controller path with transfer limits of $128$ KiB/s, $512$ KiB/s, and $2$ MiB/s. A controller-side monotonic clock measures complete response delivery. TSR counts protocol-complete outputs delivered before their deadlines. Task correctness is checked separately against the expected arithmetic results or issue metadata.

The real-world testbed uses a world model with twelve lognormal duration predictors and twelve protocol-success probabilities, covering different workers, operation types, and resource profiles. These components are fitted from independent calibration observations to capture variation in execution time and protocol completion. Together with the execution rules, they support the prediction of workflow progress and recoverable state under candidate management actions.

\subsubsection{Performance Comparison}

\begin{figure}[!tb]
    \centering
    \includegraphics[width=\linewidth,trim=0bp 13.37bp 0bp 7.02bp,clip]{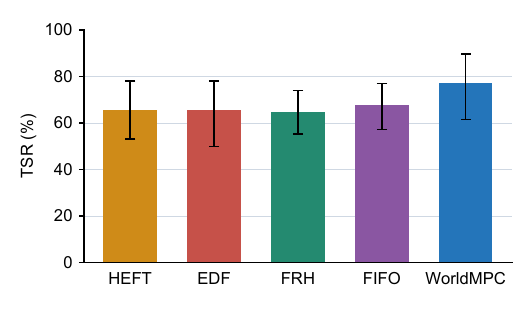}
    \caption{Task success rate on the real-world testbed.}
    \label{fig:eval_testbed_tsr}
\end{figure}

Fig.~\ref{fig:eval_testbed_tsr} compares the TSR of the five methods on the real-world testbed. WorldMPC achieves $77.08\%$, compared with $65.63\%$, $65.63\%$, $64.58\%$, and $67.71\%$ for HEFT, EDF, FRH, and FIFO, respectively. WorldMPC considers resource contention and recovery costs when selecting execution actions. This helps more requests finish within their deadlines.

\section{Conclusion}\label{sec:conclusion}

In this work, we address reliability-aware AI agent execution in collaborative edge computing under node failures and wireless disruptions. We propose a world model-based MPC approach to jointly optimize resource allocation, state protection, and recovery for evolving agent workflows with interdependent reasoning and tool progress under task deadlines. Experimental results demonstrate that the proposed approach improves task success rates over the evaluated baselines. This work contributes to reliable completion of stateful agent tasks in dynamic edge environments. Future research will focus on the generalization across diverse agent workloads and heterogeneous devices, and varying network and execution conditions.

\FloatBarrier
\bibliographystyle{ieeetr}
\bibliography{citations}

\raggedbottom
\vspace{-2em}
\begin{IEEEbiography}[{\includegraphics[width=1in,height=1.25in,clip,keepaspectratio]{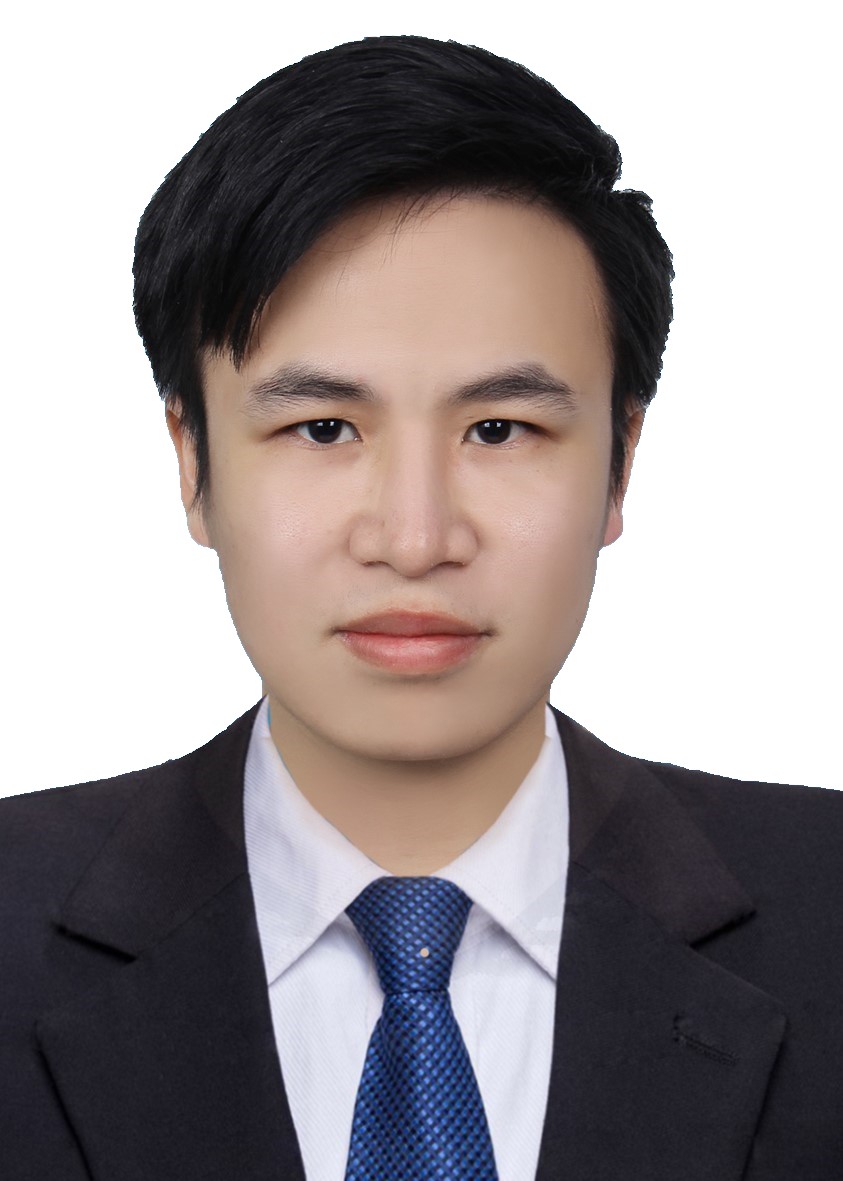}}]{Xiangchun Chen} is currently a Postdoctoral Fellow with the Department of Computing, The Hong Kong Polytechnic University, Hong Kong SAR, China. He received the B.Eng. degree in computer science and technology from Harbin Institute of Technology, China, in 2021, and the Ph.D. degree in computing from The Hong Kong Polytechnic University, Hong Kong SAR, China, in 2026. His research interests include edge AI agents, edge LLMs, reinforcement learning, and digital twins.
\end{IEEEbiography}
\vspace{-2em}

\begin{IEEEbiography}[{\includegraphics[width=1in,height=1.25in,clip,keepaspectratio,trim=60bp 163.44bp 48.96bp 57.6bp]{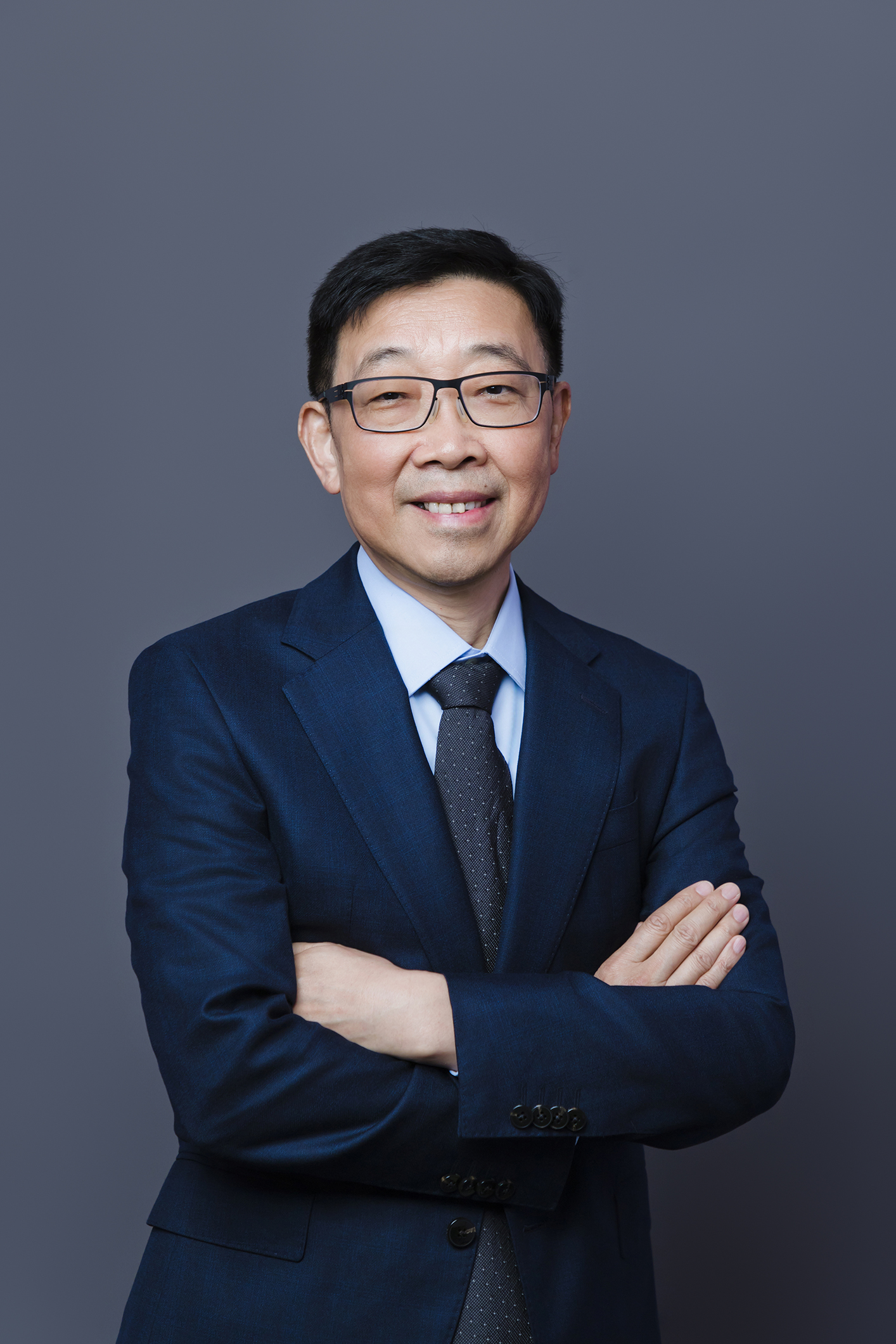}}]{Jiannong Cao (Fellow, IEEE)} is currently the Otto Poon Charitable Foundation Professor in Data Science and the Chair Professor of Distributed and Mobile Computing in the Department of Computing at The Hong Kong Polytechnic University (PolyU), Hong Kong. He is the founding director of the Research Institute for Artificial Intelligence of Things in PolyU, and the director of the Internet and Mobile Computing Lab. He is also the Vice President (Education) of PolyU. He served as the department head from 2011 to 2017. Prof. Cao is a member of Academia Europaea, a fellow of IEEE, a fellow of the China Computer Federation (CCF), and an ACM distinguished member. His research interests include distributed systems and blockchain, wireless sensing and networking, big data and machine learning, and mobile cloud and edge computing.
\end{IEEEbiography}
\vspace{-2em}

\begin{IEEEbiography}[{\includegraphics[width=1in,height=1.25in,clip,keepaspectratio,trim=0bp 16bp 0bp 0bp]{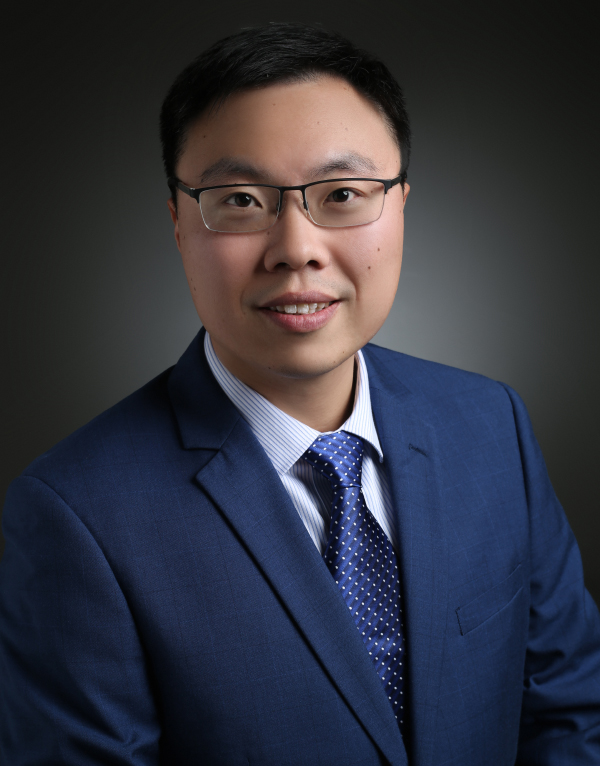}}]{Longxiang Gao} received the Ph.D. degree in computer science from Deakin University, Australia, in 2014. He is currently a Professor with the Shandong Computer Science Center, Qilu University of Technology (Shandong Academy of Sciences), Jinan, China. He was a Postdoctoral Research Fellow with IBM Research and Development, Australia, and a Lecturer and Senior Lecturer with Deakin University. His research interests include edge intelligence, network security, and quantum computing. He is a senior member of both the ACM and IEEE.
\end{IEEEbiography}
\vspace{-2em}

\begin{IEEEbiography}[{\includegraphics[width=1in,height=1.25in,clip,keepaspectratio,trim=67.5bp 247.5bp 42bp 15bp]{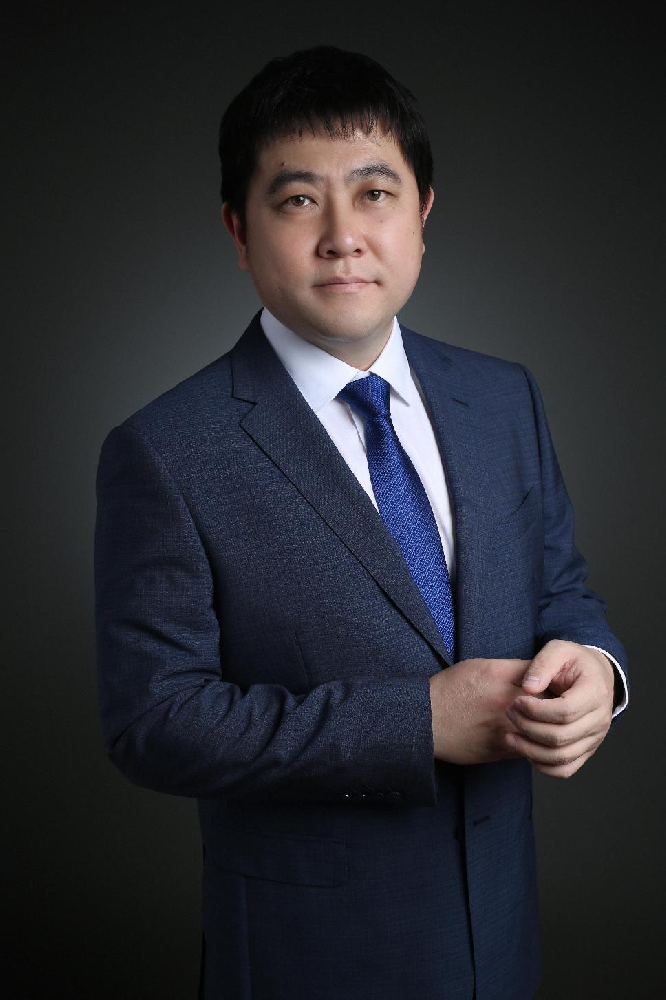}}]{Xiaoming Wu} received the M.Eng. degree in computer science and technology from Shandong University, China, in 2006, and the Ph.D. degree in software engineering from Shandong University of Science and Technology, China, in 2017. He is currently a Professor with the Shandong Computer Science Center, Qilu University of Technology (Shandong Academy of Sciences), Jinan, China. He also serves as the Director of the Faculty of Computer Science and Technology and the Shandong Computer Science Center. His research interests include cybersecurity, industrial Internet, data security, and privacy protection.
\end{IEEEbiography}
\vspace{-2em}

\begin{IEEEbiography}[{\includegraphics[width=1in,height=1.25in,clip,keepaspectratio]{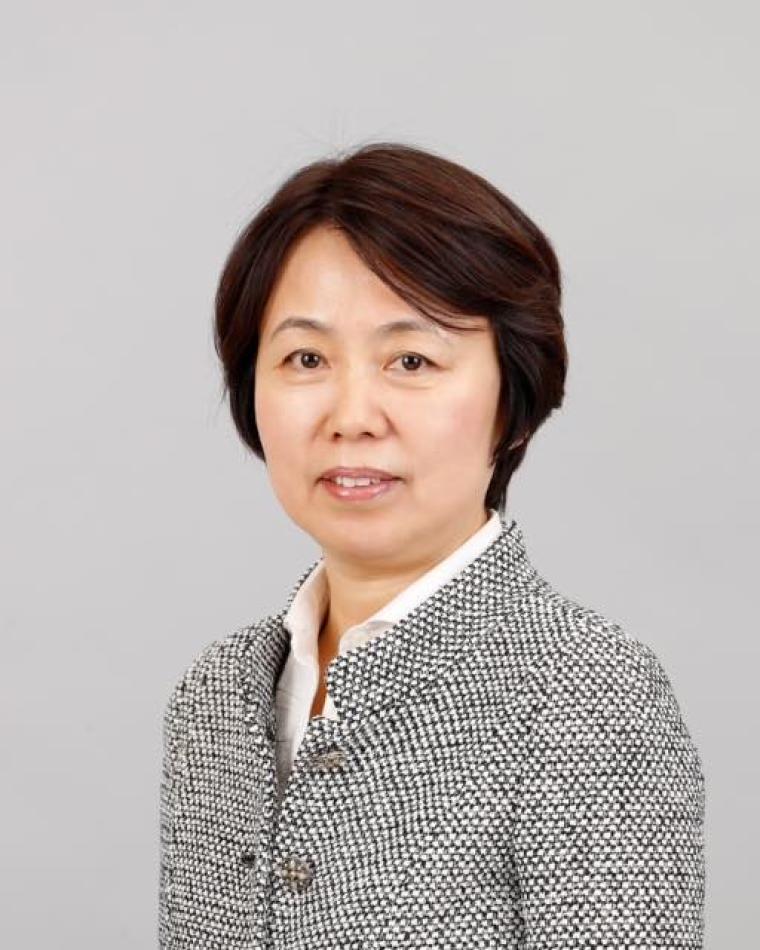}}]{Yusheng Ji (Fellow, IEEE)} received the B.E., M.E., and D.E. degrees in electrical engineering from The University of Tokyo. She joined the National Center for Science Information Systems (NACSIS), Tokyo, Japan, in 1990. She is currently a Professor Emeritus with the National Institute of Informatics (NII), Tokyo, Japan. Her research interests include network resource management and mobile computing.
\end{IEEEbiography}
\vfill

\end{document}